\pdfoutput=1
\documentclass[format=sigconf]{acmart}

\usepackage{booktabs} 
\setcopyright{rightsretained}

\copyrightyear{2018}
\acmYear{2018}
\setcopyright{acmlicensed}
\acmConference[ICSE '18]{ICSE '18: 40th International Conference on Software Engineering }{May 27-June 3, 2018}{Gothenburg, Sweden}
\acmPrice{15.00}
\acmDOI{10.1145/3180155.3180193}
\acmISBN{978-1-4503-5638-1/18/05}

\usepackage[utf8]{inputenc}
\usepackage{url}
\usepackage{enumitem}
\usepackage{flushend}
\usepackage{graphicx}
\usepackage{xspace}
\usepackage{amssymb}

\usepackage{subcaption}
\usepackage{hyperref}
\newcommand{\etal}{et al.\xspace}
\newcommand{\ie}{i.e.,\xspace}

\newcommand{\fig}[1]{Fig.~\ref{#1}}
\newcommand{\tab}[1]{Table~\ref{#1}}

\usepackage[disable]{todonotes}

\newcommand{\bram}[1]{\todo[color=orange!40, inline]{\footnotesize{Bram: #1}}}

\begin{document}
%
\title{Do Programmers Work at Night or During the Weekend?}


\author{Maëlick Claes}
\affiliation{M3S, ITEE, University of Oulu, Finland}
\email{maelick.claes@oulu.fi}

\author{Mika V. Mäntylä}
\affiliation{M3S, ITEE, University of Oulu, Finland}
\email{mika.mantyla@oulu.fi}

\author{Miikka Kuutila}
\affiliation{M3S, ITEE, University of Oulu, Finland}
\email{miikka.kuutila@oulu.fi}

\author{Bram Adams}
\affiliation{MCIS, Polytechnique Montreal, Canada}
\email{bram.adams@polymtl.ca}





\begin{abstract}
Abnormal working hours can reduce work health, general well-being, and
productivity, independent from a profession. 
To inform future approaches for automatic stress and overload detection, this paper establishes empirically
collected measures of  the work patterns of software engineers.
To this aim, we perform the first large-scale study of software engineers' working hours by investigating the
time stamps of commit activities of 86 large open source software
projects, both containing hired and volunteer developers. 
We find that two thirds of software engineers mainly follow typical
office hours, empirically established to be from 10h to 18h, and do 
not usually work during nights and weekends. 
Large variations between projects and individuals exist. Surprisingly, we found no support that project maturation would decrease abnormal working hours. 
In the Firefox case study, we found that 
hired developers work more during office hours while seniority, either in terms of number of commits or job status, did not impact working hours. 
We conclude that the use of working hours or timestamps of work products for stress detection requires establishing baselines at the level of individuals.



\end{abstract}

\maketitle


\section{Introduction}

Poor working patterns can reduce individual
health, well-being and productivity. Long working
hours have been associated with depression, anxiety, sleep
deprivation, and coronary heart disease
~\cite{bannai2014association}. A
survey study of 35,000 people showed that atypical working hours
increased health complaints and poor work life balance even after
controlling for the effect of night and shift work
~\cite{Greubel2016}. Similar results are found in other papers that
highlight health problems related to food digestion and sleeping
~\cite{Wirtz2008} as well as social impairment with respect to family
life~\cite{Giebel2008}.

Health problems of poor working patterns are perhaps caused by violation of our natural 24-hour (circadian) rhythm,
which has seen a surge of research interest in medicine. Most notably, the research establishing the genetic origins of circadian rhythm was awarded the 2017 Nobel prize in medicine ~\cite{Nobel2017}. Recent findings suggest that disturbances in circadian rhythm are the cause (not the effect) of depression and anxiety  ~\cite{timothy2017circadian}, and that artificial light at night increases the risk of obesity and cancer \cite{zubidat2017artificial}.

Fortunately, weekend recovery has been shown to
improve weekly job performance, personal initiative, organizational
citizenship behavior, and to lead to a lower perceived effort
~\cite{binnewies2010recovery}. Psychological detachment during
off-work time reduces emotional exhaustion caused by high job demands
and helps to maintain work engagement~\cite{sonnentag2010staying}. 
Although worker autonomy seems to reduce the
negative health and well-being effects of atypical working hours, it
does not completely balance them~\cite{Arlinghaus2016}. Other
studies of flexible working hours show that they provide
affective and work-life benefits~\cite{spieler2016help}, which is
compatible with the widely accepted results that worker autonomy
increases worker well-being stemming (cf. Karasek's job demands
control model)~\cite{karasek1979job}. Finally, extensive use of
flexible working time is linked with reduced worker productivity
~\cite{spieler2016help}.

In the software engineering domain, it has been observed that the bugginess of commits, i.e., negative effects on overall developer productivity, is related to the hour of the day (the so-called ``circadian work pattern'') and to day of the week those commits have been made~\cite{Kim2008}. 
However, the results between projects seem to
vary. Eyolfson \etal~\cite{Eyolfson2011,Eyolfson2014} propose, based
on three well known open source projects (Linux kernel, PostgreSQL, and
Xorg), that commits made between 00:00 and 04:00 contain more bugs,
while commits made between 07:00 and 12:00 (noon) contain the
least. Prechelt and Pepper~\cite{Prechelt2014} demonstrate, using data from a
closed source industry project, that the most defect-prone hour was
20:00. A hypothesis that unifies these two findings could be that
all of those buggy hours might in fact demonstrate the end of a working stint
where a developer just wants to be done with the task, leading to
prematurely committing the code. Finally, industrial software developers have also developed an interest 
to night work: 
\begin{itemize}
  \item Quora question with over 100 answers and over 500,000 views on why developers love night work~\cite{QuoraWhy2014} 
  \item entire book with 2,500 paid readers on the topic, written by an industrial programmer~\cite{Teller2013}
  \item Stackoverflow analysis with 3,800 Facebook shares and 91 comments of day-night differences~\cite{LanguagesSO}
\end{itemize}

As abnormal working hours can affect occupational health,
well-being, productivity and staff turnover among software engineers,
this paper aims to investigate the work patterns of software developers in large open
source projects, either projects with many hired resources (Mozilla)
and those without (Apache Foundation), as well as in a (local) Finnish
IT company. We are particularly interested in the degree to which work
is performed outside of the commonly expected working hours, since, based on the literature discussed above, such
irregular working hours may act as a proxy for job-related stress and
time pressure conditions and suggest non-sufficient detachment from
work.

The Mozilla Foundation is known for Firefox, which makes an
interesting case study as the core software engineers are paid to develop
it. In addition, Mozilla also hosts projects that have been gradually
abandoned and left to the community, such as Thunderbird and
SeaMonkey. The Apache Foundation is an organization supporting open
source projects and their communities of developers, but does not pay
its developers. Some of its projects are business-critical to
companies, who pay developers to work on them, while others consist of
hobbyist developers. Finally, the local company provides a reference
to compare the groups of open source projects to, since a closed
source company follows a more traditional work schedule and larger
financial concerns are at stake.

First, we investigate the following research questions for all
considered projects:
\begin{description}
\item[RQ1] {\bf What are the circadian and weekly work patterns of
  software developers?} 
\item[RQ2] {\bf How does the usual work pattern vary across different
  projects?} 
\item[RQ3] {\bf Are office hour commits different in terms of
  size?} 
\item[RQ4] {\bf Is there a difference in the developers' work patterns
  over time?} 
\end{description}

Then, because results of RQ5 can considerably vary due to differences
in terminology between projects, we focus solely on Mozilla
Firefox. Moreover, the identification of developers background needed
for RQ6 involves a lot of manual work and is too laborious to be
performed for all projects. Furthermore, Firefox is the largest
project of our data set as it contains 228,697 commits, which
represents roughly one third of all of the commits of our data
set. Finally, Firefox includes both paid work and voluntary
contributions.

\begin{description}
\item[RQ5] {\bf Are office hour commits different in terms of content?}
\item[RQ6] {\bf Can demographics explain office hour activity?}
\end{description}

This paper is structured as follows. We discuss related work in
Section~\ref{sec:related}. In Section~\ref{sec:data}, we give details
on our data extraction process, then we address our research questions
in Section~\ref{sec:analysis_all} and
Section~\ref{sec:analysis_ff}. We then present the threats to validity
that can impact our results in Section~\ref{sec:threats} and conclude
in Section~\ref{sec:conclusion}.



\section{Related Work}\label{sec:related}

Sall \etal~\cite{Sall2007} studied weekend work activity patterns in
the San Francisco Bay Area using surveys. Their results indicate that
a host of variables affect the likeliness of working during the
weekend, in particular, gender, race, type of work and
income. Individuals are more likely to work out of home during
weekends in the winter season than in other seasons.

Wang \etal~\cite{DBLP:journals/corr/abs-1208-2686} examined work
patterns of scientists by looking at the amount of scientific papers
being downloaded on different days. Scientists work 60-70\% as
much of their time during the weekend as during the week. Time worked
during weekends differs by country: scientists work proportionally
more during weekends in China than in the USA and Germany.

Binnewies \etal~\cite{binnewies2010recovery} investigated the
importance of recovery during weekends and its implications on work
performance. Data from surveys indicate that experiencing
psychological detachment, relaxation and mastery during weekends was
positively correlated with being recovered at the beginning of the
working week, which in turn was positively related to self-reported
work performance.

McKee~\cite{McKeei2750} investigated the reasons for increased mortality
rate in hospitals during the weekend, with explanations ranging from
more seriously ill patients to less experienced staff. In extreme cases,
the weekend effect to lead to 44\% higher odds of
mortality on Friday compared to Monday~\cite{Aylinh4652}. However,
multiple sources state conflicting evidence on the source of this
effect~\cite{Alexander2010,Aylinh4652,McKeei2750}.


Some relevant studies also exist in software engineering. Industrial
blogs on GitHub~\cite{LanguagesGH} and
StackOverflow~\cite{LanguagesSO} report that less main stream
languages such as Haskell are more commonly used during the night than
languages adopted in the industry such as Java. Multiple
studies~\cite{Ferrucci2013, Barros2016, Sarro2017} proposed
multi-objective techniques to support project planning avoiding overtime.

Khomh \etal~\cite{Khomh2015} studied the impact of Firefox's fast
release cycle on post-release bugs and found that not only did the
new release cycle not increase the number of bugs, bugs were also
fixed faster. Although most of the studies thus far on the switch of
release cycle have focused on its quality assurance implications, the
repercussions for developers in terms of work quality have largely
been ignored.

Our previous work~\cite{claes2017abnormal} studied 
abnormal working hours on two projects only, while this paper performs a large-scale study on 86 open source and 1 industrial project. Furthermore,
we use more advanced methods: k-means clustering and 
a dynamic search of office hours instead of a static heuristic. 

To summarize, although some initial evidence has been found regarding
the interplay between work and well-being in software engineering, a
structured analysis of (un)healthy work patterns is missing. This
paper starts to fill this gap by studying and comparing the periods
during which developers are actively working in a software project.




\section{Data extraction}\label{sec:data}

We
mined development data from the Git and Mercurial repositories of
Mozilla\footnote{\url{https://hg.mozilla.org}} and
Apache\footnote{https://git.apache.org/} and of a local company's product. The local company's product contained more than 20,000 commits from nine developers.  We have visited the company several times to ensure the validity and usefulness of our work. We then wrote custom scripts to extract the list of commits (code
changes), associated timestamps and authors from all code
repositories.

For Mozilla, we needed to do extra processing in order to find out which commit belongs to which project. 
First, we used the \emph{GrimoireLab}
tools\footnote{\url{https://grimoirelab.github.io/}} to extract issue
comments from Mozilla's Bugzilla
repository\footnote{\url{http://bugzilla.mozilla.com}} (i.e., the
database containing reported issues, such as bug reports or feature
requests). 
Second, we linked commit messages to the
corresponding issue report by looking for an issue identifier in a
given message. Out of the 396,180 extracted commits, 330,078
  were successfully linked to a bug issue. After linking, we then
filtered the commits to only keep the ones related to the following
major \emph{products}: \emph{Firefox}, \emph{Core}, \emph{Firefox OS},
\emph{Firefox for Android}, \emph{Thunderbird} and \emph{SeaMonkey}.

While the history of each Apache project is stored in an individual
git repository, we realized that many of these old commits were
missing the time zone information. These commits might have been
imported from another version control system that did not store such
information. Thus, for each project, we only considered commits starting 
from 
the first commit with a timezone different than
UTC (i.e., the first commit that is clearly made after the import from the old
system). Moreover, we only considered Apache projects with at least
2,000 commits after filtering, which reduced the number of Apache projects
from 822 to 81. This left us with a total of 451,116 commits.


In order to study the work patterns of individual developers,
we performed a basic merging of the different authors' identities. We
first cleaned the name and email used in the version control system's
author field. Then we grouped together identities using the same name
or email addresses. Finally, two of the authors manually checked the
result in order to avoid any false positive.


\section{Empirical analysis of the work patterns of all projects}\label{sec:analysis_all}

\subsection*{RQ1. What are the circadian and weekly work patterns of software developers?}

\begin{figure*}[t]
  \centering
  \begin{subfigure}{0.48\textwidth}
    \centering
    \includegraphics[width=\textwidth]{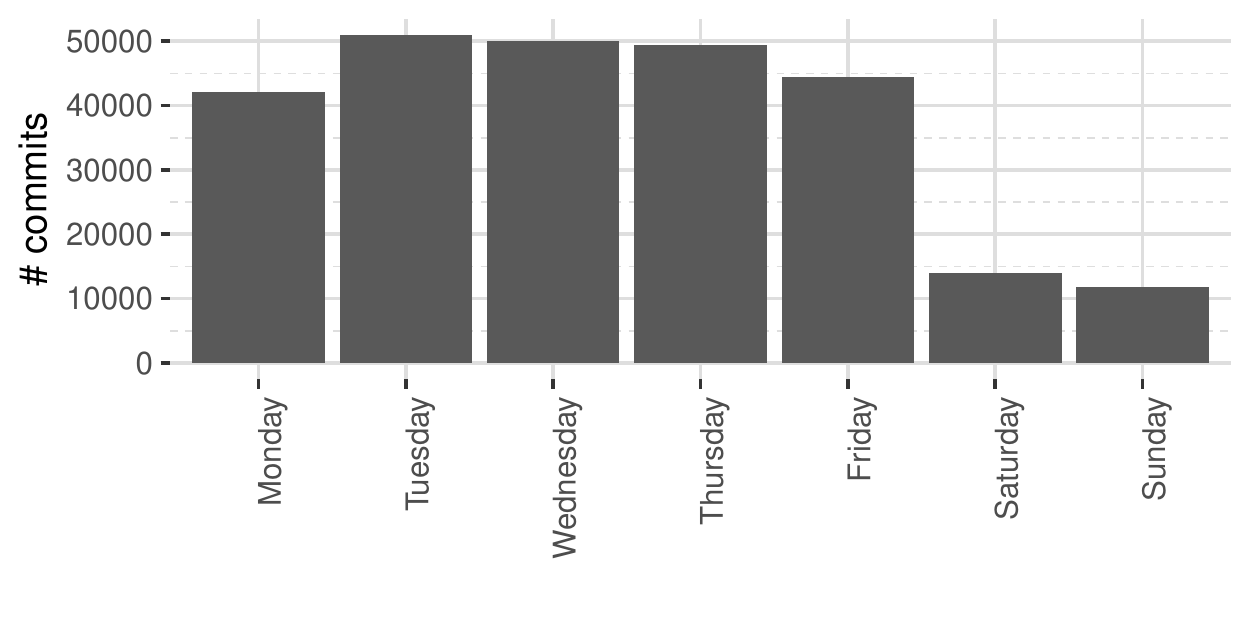}
    \caption{Daily for all Mozilla projects}
    \label{fig:mozilla-weekdays}
  \end{subfigure}
  \begin{subfigure}{0.48\textwidth}
    \centering
    \includegraphics[width=\textwidth]{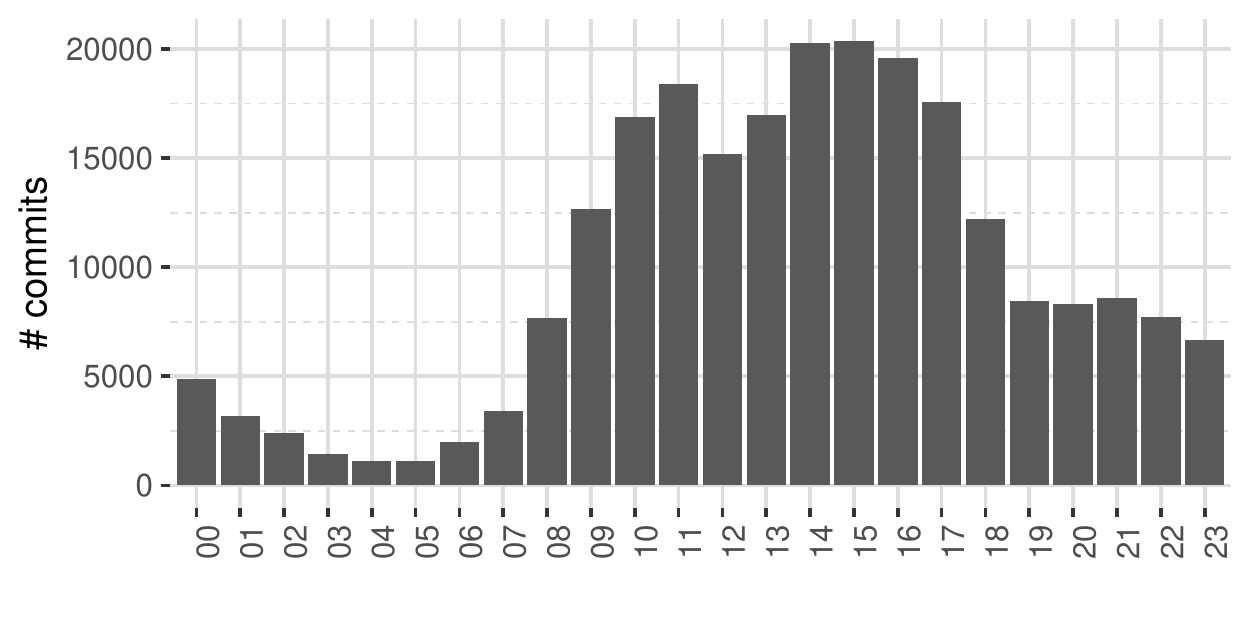}
    \caption{Hourly for all Mozilla projects}
    \label{fig:mozilla-hours}
  \end{subfigure}

  \begin{subfigure}{0.48\textwidth}
    \centering
    \includegraphics[width=\textwidth]{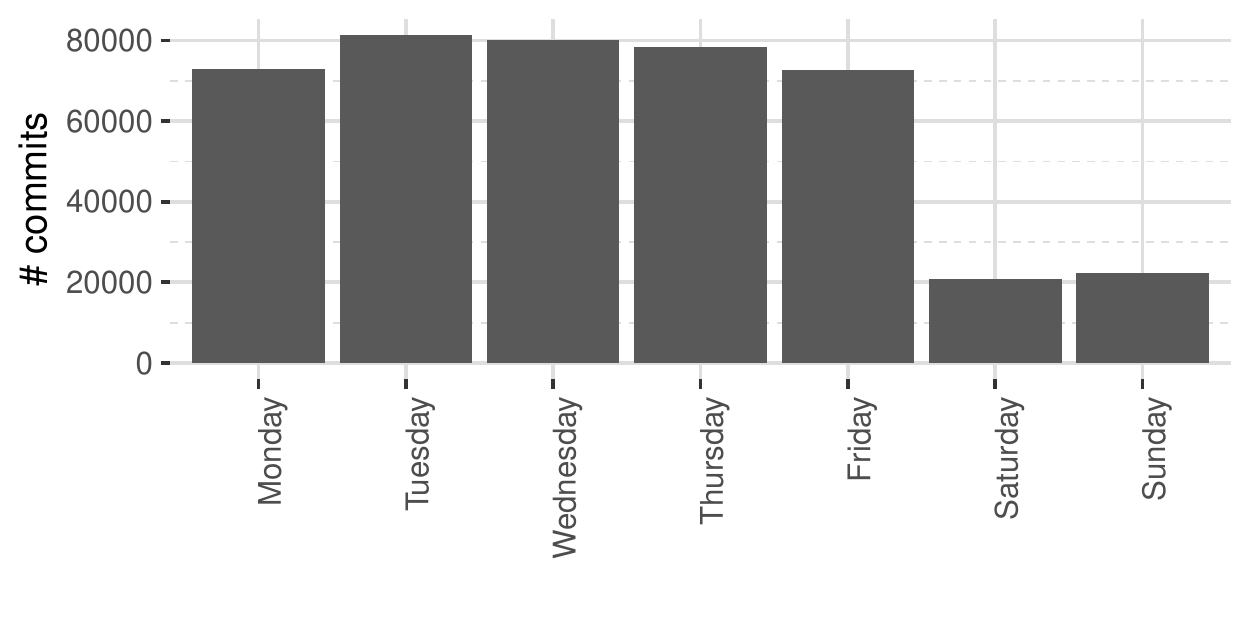}
    \caption{Daily for all Apache projects}
    \label{fig:apache-weekdays}
  \end{subfigure}
  \begin{subfigure}{0.48\textwidth}
    \centering
    \includegraphics[width=\textwidth]{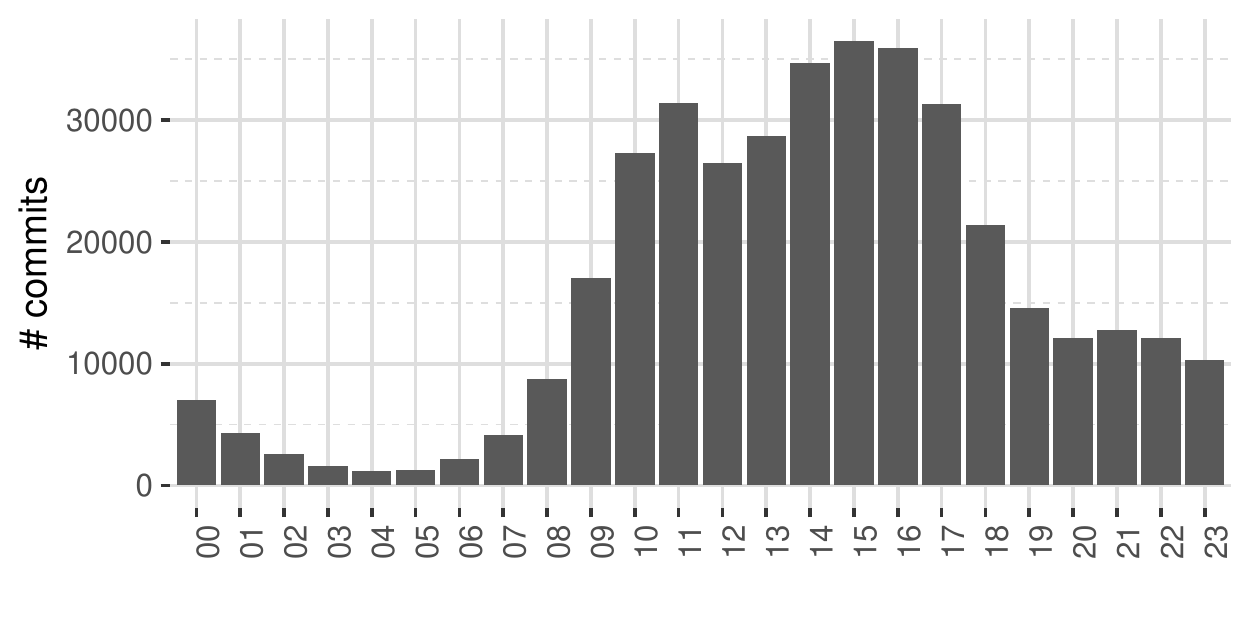}
    \caption{Hourly for all Apache projects}
    \label{fig:apache-hours}
  \end{subfigure}

  \begin{subfigure}{0.48\textwidth}
    \centering
    \includegraphics[width=\textwidth]{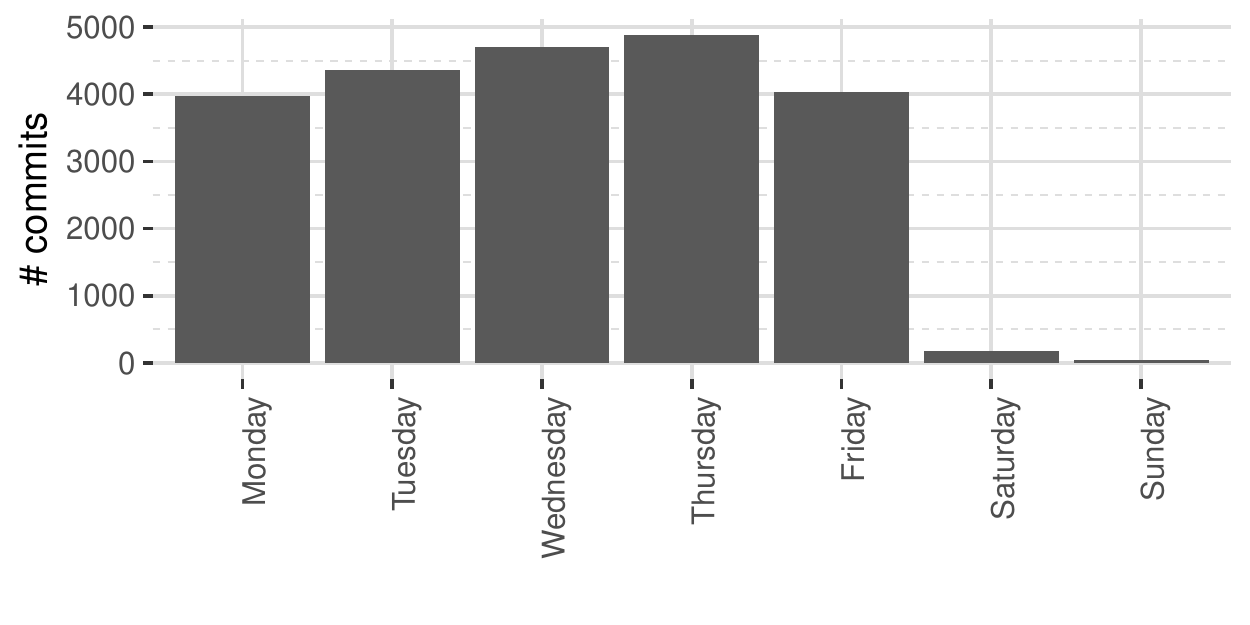}
    \caption{Daily for the local company}
    \label{fig:local-weekdays}
  \end{subfigure}
  \begin{subfigure}{0.48\textwidth}
    \centering
    \includegraphics[width=\textwidth]{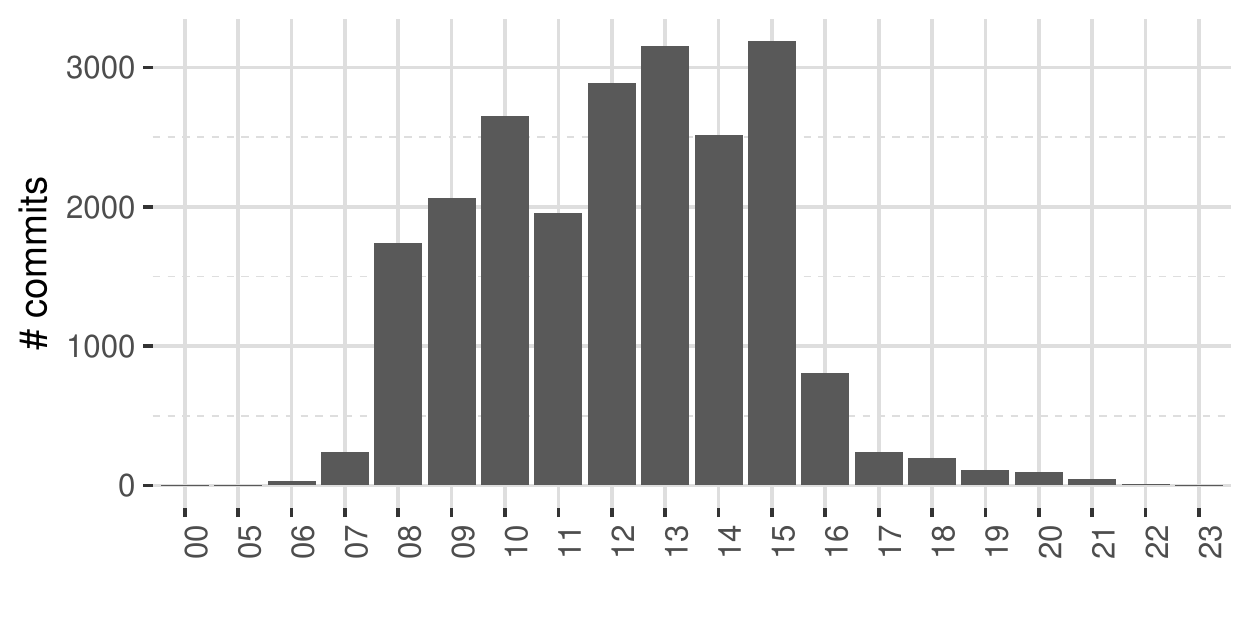}
    \caption{Hourly for the local company}
    \label{fig:local-hours}
  \end{subfigure}

  \caption{Distribution of the number of commits across the days in a week and the different
    hours in a weekday.}
\end{figure*}


\textbf{Motivation.} Given the well-being and health risks related to
abnormal working hours, we wish to establish empirical baselines that
later may be turned into normative guidelines to ensure the well-being
and health of software developers.

\textbf{Commits follow a weekly rhythm.}
\fig{fig:mozilla-weekdays},~\ref{fig:apache-weekdays}
and~\ref{fig:local-weekdays} show the number of commits made on each
day of the week within the Mozilla (319,139 considered commits),
Apache (574,563 commits) and the local company projects (22,193 commits),
respectively. We can clearly see that fewer commits are being posted
during the weekend, while there is a slight variation in activity
during the week, with Monday and Friday being the least active days in
all cases. Tuesday is the most active day for the Mozilla and Apache
projects, while in the local company it is Thursday.

\textbf{Commits follow a circadian (24-hour) rhythm.}  The
circadian rhythm is present in Apache, Mozilla and in the local
company projects, see \fig{fig:mozilla-hours},~\ref{fig:apache-hours}
and~\ref{fig:local-hours}. The lowest activity for all projects
happens during the night, with the number of commits starting to increase in the
morning until a dip in activity happens during
lunch hour. In the Mozilla and Apache projects, lunch mostly happens at 12 o'clock,
while in the local company it happens at 11 o'clock (something which we
confirmed through face-to-face discussions with the
company).

After lunch hour, the number of commits increases again until a
decreasing trend in commits starts setting in from 4pm in all
 three data sets. For the local company, we can also see an
unexpected dip in commits at 2pm, yet the company
was not able to explain this phenomenon to us. The
decreasing trend in the number of commits plateaus at 7pm
for the Apache and Mozilla projects, while for the local company, we can see
that work during the evenings is very limited compared with the
Mozilla and Apache projects.

\textbf{Working hours of software developers are typically from
  10-18.}  If we consider a 40 hour work week, which is still the norm
for most professions, each week day should correspond to about 8
working hours. Hence, in order to determine the working period for
each of our projects, we searched the week days for the eight-hour
stint that covers the largest share of commits in a day. First, for a
given project, we computed all the $(HH,mm)$ tuples representing a timestamp
(in 24 hours format) for which at least one commit was made during a
weekday. For each $(HH,mm)$ tuple, we computed the number of commits made
during the 8-hour interval $[(HH,mm),(HH_{+8},mm)[$. We then selected the
interval with the highest number of commits as the 8-hour office
period of the considered project.

The results of this search are shown in Table~\ref{tab:workinghours}. For 34\% (30/87) of the projects, this
search results in working hours starting between 09:00 and 09:59, and for
45\% (39/87) of the projects, the working hours start between 10:00 and
10:59. The median start time of work across all projects is at 10:03.

\begin{table*}[]
\centering
\caption{Start of working hours for projects in our data set.}
\label{tab:workinghours}
\begin{tabular}{l|l|l|l|l|l|l|l|l|l}
  Start time & 08:00-59 & 09:00-59 & 10:00-59 & 11:00-59 & 12:00-59 & 14:00-59 & 16:00-59 & 17:00-59 \\
  \hline
  \# Projects & 4 & 30 & 39 & 8 & 2 & 1 & 2 & 1
\end{tabular}
\end{table*}


\textbf{66\% of developers follow office hours.} 
We were
interested in seeing if we can find different work patterns across the
weekly cycle and the daily circadian rhythm of different developers within our data set of 87
software projects, and how those patterns look like. While past work has
considered the core contributors to be the top 20\% of the
committers~\cite {robles2006contributor}, in our data set it only takes 22 commits total to be in the top
, which we think is not
enough for establishing weekly working patterns. Thus we only
consider the top 10\% of developers, leaving us with 1,108 developers
(out of 11,059), each with 95 or more commits. This top 10\% of
developers has made 88\% out of all commits of our data set.

\begin{figure}[t]
  \centering
  \includegraphics[width=0.8\columnwidth]{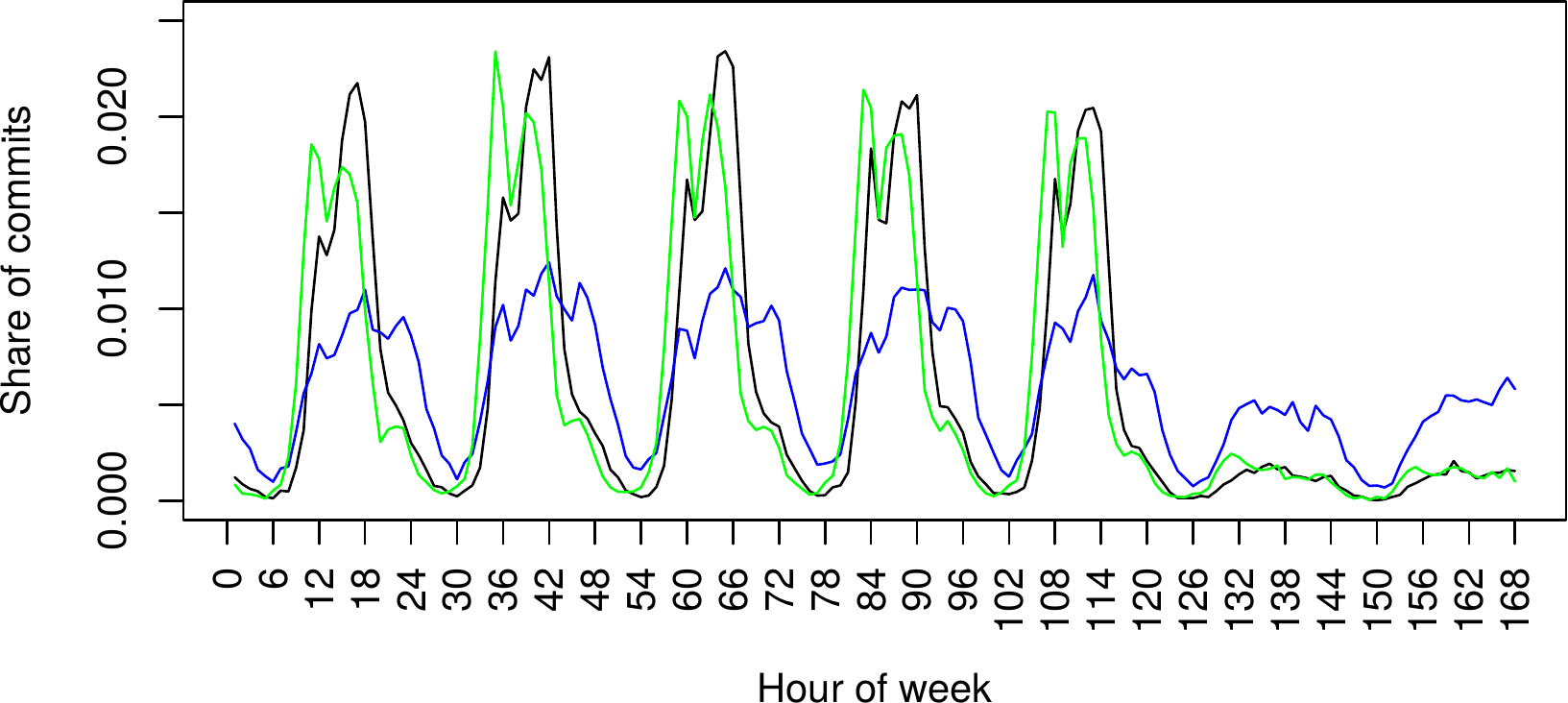}
  \caption{Weekly work patterns for the three k-means centroids corresponding to the 1,108 top software developers.}
  \label{fig:All_weekly_hour_cluster}
\end{figure}

For clustering purposes, we computed the relative share of commits made
by each hour of the week by each developer. Thus, for each developer, we end up with a
feature vector of 168 elements, one for each hour of the week. 
Similar to previous work on 
weekly patterns of mobile phone usage behavior
~\cite{csaji2013exploring}, we then used k-means clustering on the vectors. We tested values of k (i.e., the number of clusters) from 2 to 7. We observed that going beyond three clusters did not bring new information, instead clusters with very few individuals and high noise started to appear.  
\fig{fig:All_weekly_hour_cluster} shows the weekly work patterns (the feature vectors of centroids) of each of the three
clusters.

The figure shows that two clusters (green and black lines)
follow typical office hours where commits are concentrated on day
time hours during the work week, with clear
dips during lunch hour. The green cluster starts and stops working a bit earlier
than the black cluster. In fact,
when we shift the green cluster by one hour, we can see in Figure
~\ref{fig:All_weekly_hour_cluster} that both clusters have highly
similar work patterns. These two clusters contain 66\% of all
developers, and they roughly correspond to the projects starting at 9am vs. 10am in Table ~\ref{tab:workinghours}.

However, \fig{fig:All_weekly_hour_cluster} also shows a
third cluster (blue) that does not follow the regular office hour
pattern. Developers in this blue cluster commit mostly from noon until
midnight during the weekdays and weekends. The difference between the
blue cluster and the other two clusters is that this group works more
during evenings and weekends than the other two clusters.

\textbf{Discussion:} Up until now, we have established that a weekly and
circadian rhythm is followed in the projects of Apache and Mozilla
foundation, as well in the local company product that was used as a
comparison. We have also established a way to determine the eight-hour
work day start and end times for each project, then
presented measurements of how much work gets done during the typical
working hours. Establishing the normal working hours is important (especially in a distributed context) if we wish to
determine whether the developers of a particular project are, for
example, under time pressure and forced to work overtime. 

Committing 70\% of work during typical working hours might be a stress
signal for a project that has previously had a share of 90\%, but not
for another project that constantly commits 70\% work during typical
working hours. We also found that two thirds of the developers are
clustered to groups that follow office hours. In the two office hour
groups, we saw a phase shift of a bit more than one hour
. We think that this phase shift
is due to both project culture and individual preferences, as some
individuals are known as morning persons while others are
not~\cite{hu2016gwas}. This suggests that clustering of individual
developers may allow personalized stress or overtime detection, for at
least two thirds of the developers that follow regular rhythm. 

\subsection*{RQ2. How does the usual work pattern vary across different projects?}
\textbf{Motivation:} Here, we investigate variations in working hours between projects. After all, work patterns are shaped by both individual preferences and project culture and norms. 
\begin{figure}[t]
  \centering
  \includegraphics[width=0.7\columnwidth]{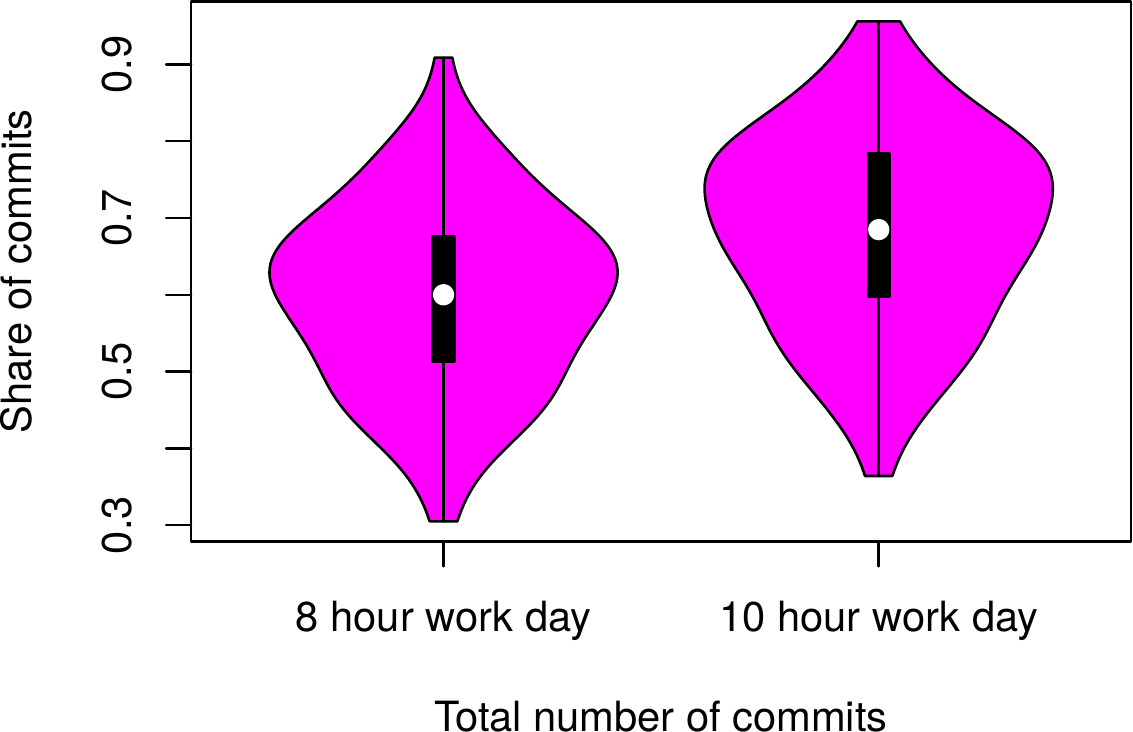}
  \caption{Distribution of percentage of commits across projects in an eight-hour typical work day and
    ten-hour extended work day (eight-hour day plus/minus one) in our
    data set of 87 projects.}
  \label{fig:violin_commits}
\end{figure}

\begin{figure}[t]
  \centering
  \includegraphics[width=\columnwidth]{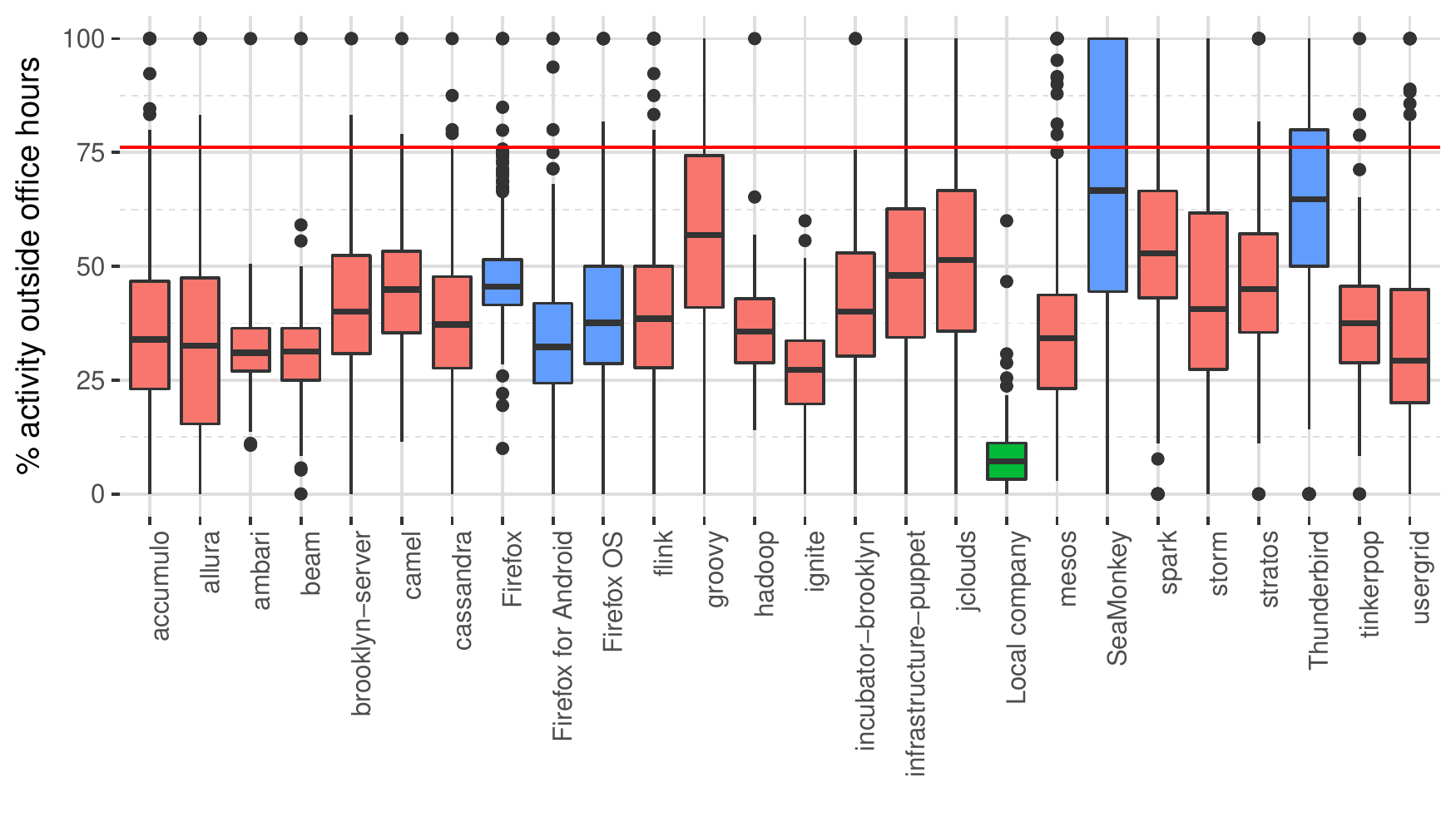}
  \caption{Variation within the local company project (green), the
    Mozilla projects (blue) and the top-20 Apache projects (red) of
    the distribution across commits done outside typical working hours
    per week.}
  \label{fig:boxplot-abnormal}
\end{figure}

\textbf{90\% of the projects have similar typical working hours.}
This can be observed from \tab{tab:workinghours}, which shows how 78
out of 87 projects work from 10-18 plus/minus one
hour. \textbf{Variations in work patterns between projects are still
  considerable} as the amount of commits done in eight-hour work days
is 60\% on average with a standard deviation of 12.0\% (min-max
31-91\%). Moreover, the extended typical hours (8 hours plus/minus one
hour) has an average of 68\% with a standard deviation of 12.6\%
(mix-max: 36-96\%).
\textbf{On average, only 60\% of work gets done during typical working
  hours.}
The violin plot in \fig{fig:violin_commits} shows
the distribution across projects of the percentage of commits
performed during working hours.

Figure~\ref{fig:boxplot-abnormal} shows the weekly variation within
the most active projects. The differences between projects in terms of
relative activity happening during 8-hour work days can partially be
explained by the presence of paid developers. In particular, the three
projects with the most activity outside office hours (SeaMonkey,
Thunderbird and Groovy) are projects that are now community projects.


\textbf{The amount of activity outside office hours during the week
  and during the weekend are correlated.} The percentage of activity
in a project outside office hours during weekdays, and the activity
during weekends yield a correlation coefficient of 0.77 (pearson),
with a p-value of $< 0.001$. When considering developers (with at
least 100 commits) instead of projects, we obtain a correlation
coefficient of 0.62. This means that both projects and developers that
work a lot outside office hours are also likely to work a lot during
the weekend.

\begin{figure}[t]
  \centering
  \includegraphics[width=0.8\columnwidth]{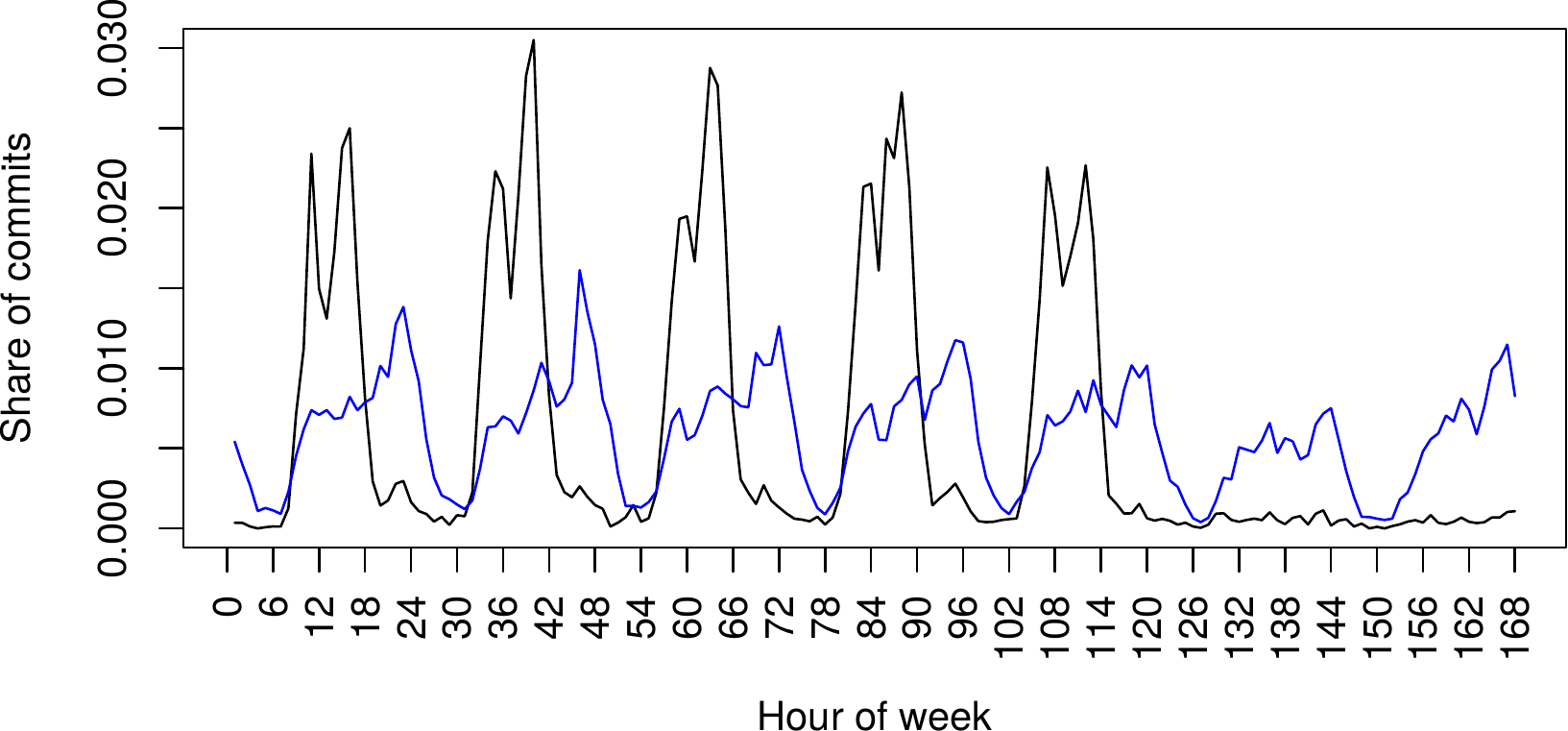}
  \caption{Two extreme clusters with respect to working hours with
    five (Black) and six (Blue) projects respectively.}
  \label{fig:projects-2-extreme-clusters}
\end{figure}

\textbf{Clustering allows the separation of projects based on work
  patterns.} We again used k-means clustering (with $k=7$) by the hour
of the week. The difference is that here we cluster by project (n=87)
while previously we clustered the top 10\% individuals (n=1,108). 
\fig{fig:projects-2-extreme-clusters} shows the two most extreme
clusters: the black one follows office hours rigorously, while the blue
one has a commit peak right before midnight. In the rigorous office
hour group (5 projects) we have our local company but also open source
projects like Apache Cordova for Android and Apache Geode. The other
group (6 projects) contains projects like SeaMonkey and Thunderbird, which are community projects (\ie very little paid work from the
Mozilla foundation) and also the logging library Log4j2. The other
clusters, not shown in the figure, fall between these two extreme
clusters.

\textbf{Discussion:} Important differences between projects can be
observed in terms of activity outside regular office hours. Although
no open source project follows office hours as well as the local
Finnish company, some open source projects follow them more closely
than others. In particular, community projects such as SeaMonkey,
Thunderbird and Groovy tend to contain a lot of activity outside
office hours.

\subsection*{RQ3. Are office hour commits different in terms of size?}

\textbf{Motivation:} Given that developers are more active during office hours than during the other hours
of the week, we wanted to check whether the commits made during
office hours are different in terms of size and content in comparison
to outside-office hour comments. Previous work has shown that
different programming languages are used during the
night~\cite{LanguagesGH,LanguagesSO} and that 
night work is
more technical~\cite{claes2017abnormal}. In this paper, we approximate the
dichotomy of office hours versus non-office hours using our dynamic office
hour detection approach, which ensures that, for each project, only the most active weekday hours are taken into account as office time.

\textbf{We found no significant difference 
in terms of added lines of code}, with a Mann–Whitney U test p-value of 0.076,
Cliff’s delta $< 0.00$. Since our projects are of different sizes, and since
a large project like Firefox may mask effects present in smaller
projects, we checked each project individually and found that for 16\%
of them, there is no difference in the median commit size. For 33\% (29/87),
the office hour commits are larger but only for 11\% (10/87) this difference is statistically significant with Mann–Whitney U test alpha
level 0.05. On the other hand, for 51\% (44/87) of the projects, the office hour
commits are smaller, while only for 26\% (23/87) of the projects a statistically significant difference was obtained. We checked the Cliffs’ Delta
effect sizes for all projects and they ranged from 0.10 to -0.11,
meaning a negligible effect in all cases~\cite{romano2006appropriate}.

\textbf{For lines of code removed, we found a significant difference
(Mann–Whitney U test p-value $< 0.001$), but the effect size is negligible
(Cliff’s delta $< 0.00$)}. Checking each project individually shows that
for 26\% (23/87) the office hour removals are larger (8\% (7/87)
significant), for 44\% (38/87) they are smaller (21\% (18/87)
significant), and for 30\% (26/87) of the projects there is no difference
in the median number of lines removed. The effect sizes were again
negligible ranging from 0.08 to -0.10.

\textbf{Discussion:} In terms of commit size, we conclude that there is no difference in practice between office hours and outside office hours.
\subsection*{RQ4. Is there a difference in the developers' work patterns
  over time?}

\textbf{Motivation:} While a software project could start off on schedule, with most of the work happening during office hours, deadlines and delays are notorious for introducing outside office work or even death marches. 
On the other hand, we could hypothesize that a project may initially have a start-up culture with highly flexible working hours, while later, as the project matures, more work would start to happen during office hours. We are interested which of these conflicting ideas, i.e., 1) starting on time with eventual delays ending up in death march versus 2) starting with a start-up culture and maturing as time passes, is more prevalent in our projects. 

\begin{figure}[t]
  \centering
  \includegraphics[width=0.8\columnwidth]{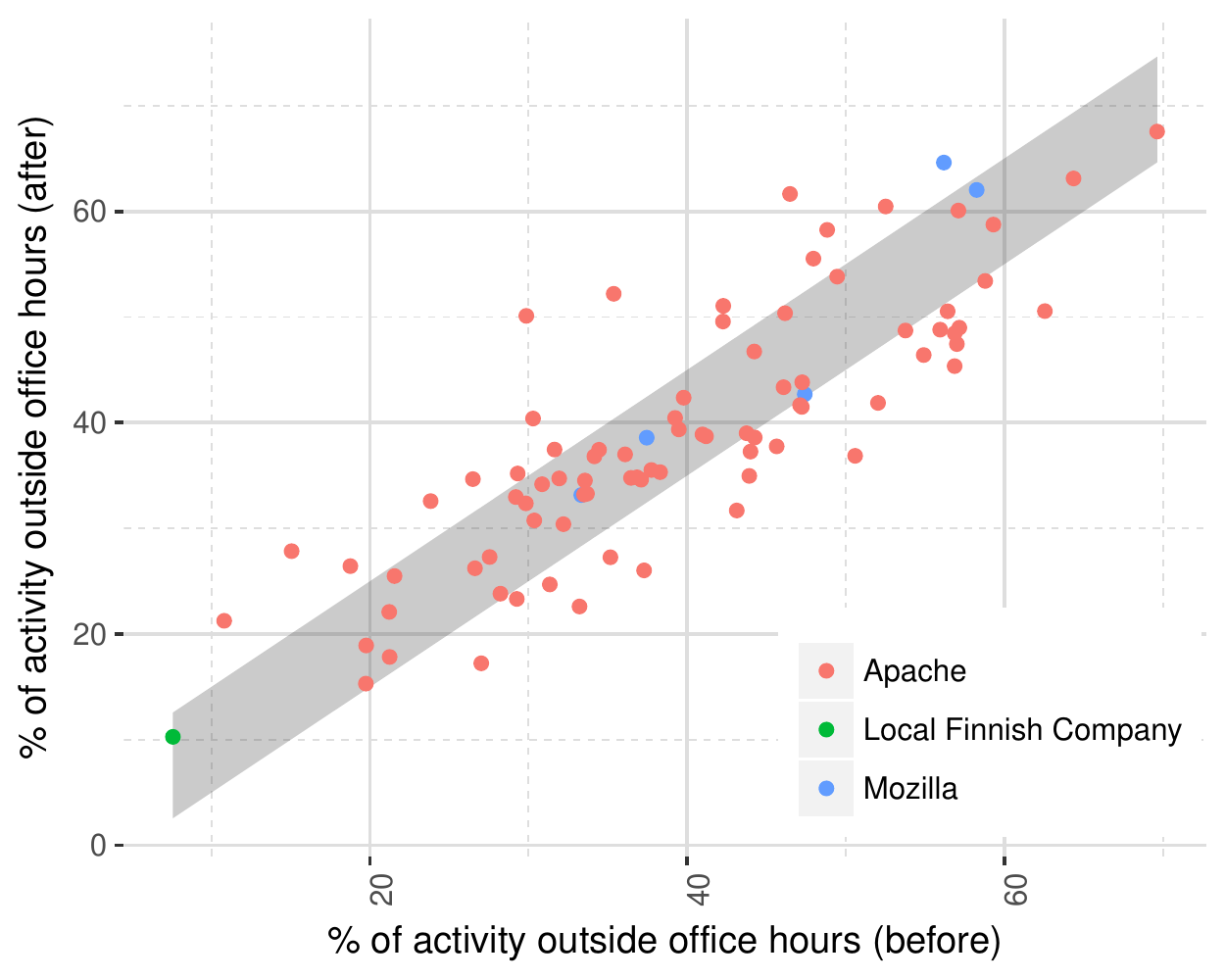}
  \caption{Percentage of activity outside office hours during the
    first and second half of each project's analyzed time
    period. Projects in the gray area did not experience an increase or
    decrease of activity of more than 5\% outside office hours.}
  \label{fig:scatter-before-after}
\end{figure}

\textbf{For 18 projects, the activity outside office hours increased by
  at least 5\%, while for 27 projects it decreased by at least 5\%.} 
For each project, we split 
the activity history in two equal parts (in terms of number of
commits) in order to find whether the work pattern changed over time
or not. \fig{fig:scatter-before-after} shows that 
for 46 projects the activity outside office hours activity neither
increased or decreased by more than 5\%. 

Examples of Apache projects where activity outside office hours
increased the most are Allura, from 26.5\% to 34.7\%, Groovy from 48.8
to 58.2\%, and Wicket from 29.8 to 50.1\%. Apache projects that saw
their activity drop include Drill from 37 to 24.4\%, AsterixDB from
57\% to 47.4\%, Spark from 54.9 to 46.4\% and Cassandra from 43.1 to
31.7\%. For Mozilla, SeaMonkey experienced a significantly increased
activity outside office hours, from 56.2\% to 64.6\% and Thunderbird
from 58.2\% to 62\%. For Mozilla, only Firefox experienced a significant decrease
from 47.4\% to 42.7\%.

\begin{figure}[!htpb]
  \centering
  \includegraphics[width=0.8\columnwidth]{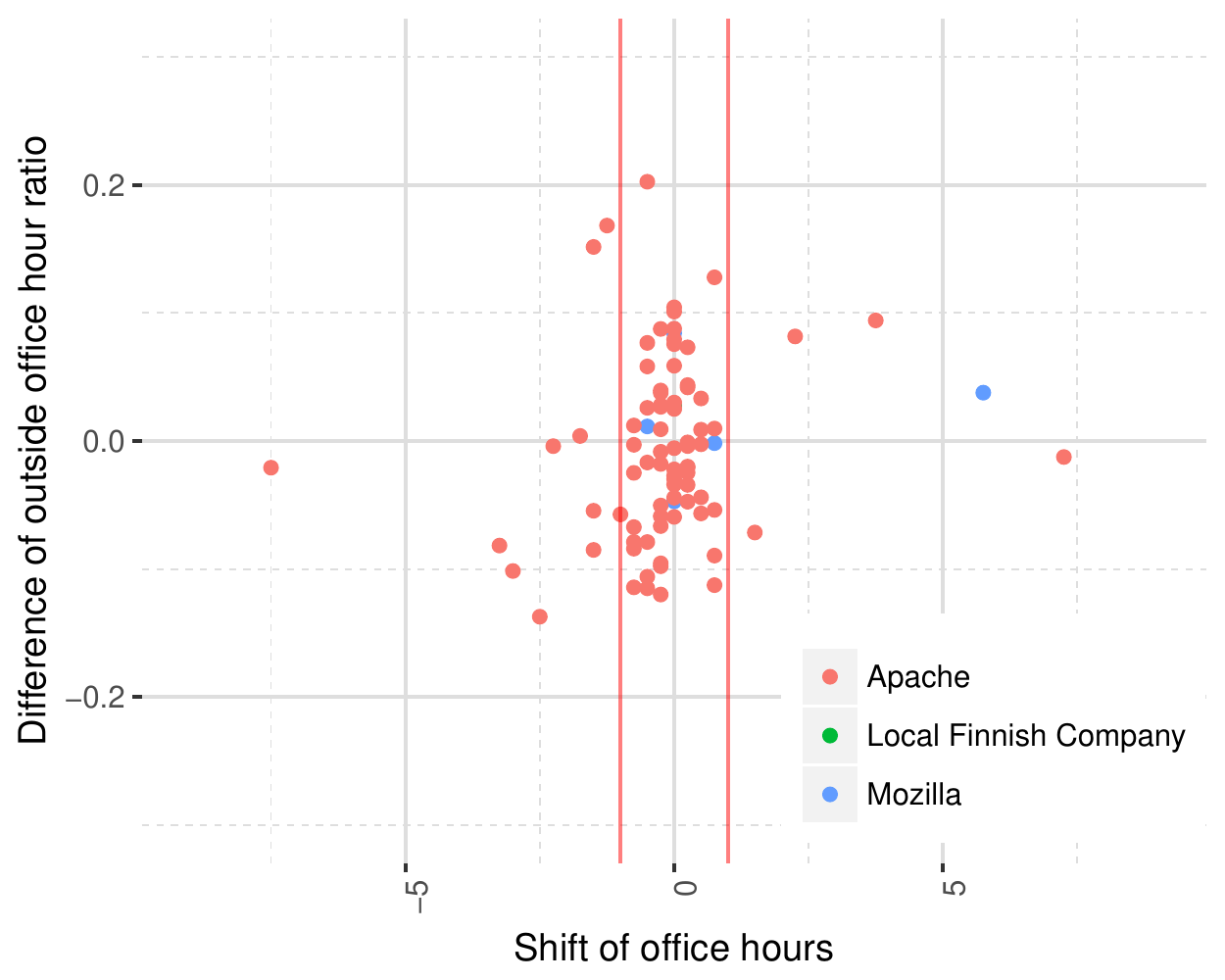}
  \caption{Differences of relative activity outside office hours over
    time (Y axis), as observed in \fig{fig:scatter-before-after}, and shifts of office hour period over time (X
    axis).}
  \label{fig:scatter-dynamic-diff}
\end{figure}

In \fig{fig:scatter-before-after}, we identified separate office hours
for each half of the projects' history, since the eight-hour office
hour time interval can change over time. As shown in
\fig{fig:scatter-dynamic-diff}, most projects do not encounter a major
shift of their time interval. Indeed, most projects (75/91)
experienced a shift of less than one hour (X axis). Only 5 projects
experienced a positive shift (towards the evening) of more than one
hour, and 11 a negative shift (towards the morning) of more than one
hour.

A significant negative shift usually means that the identified 8-hour
period shifts shift towards common office hours. For example, Apache
Log4j shifted from an 8-hour period starting at 17:15 to one starting
at 09:45. On the other hand, a positive shift usually means that the
8-hour period moves away from common office hours. For example,
Mozilla Thunderbird experienced a shift from 08:45 to 14:30. However,
in both of these cases the amount of work done outside these 8-hour
periods didn't change a lot and is relatively high (from 0.58 to 0.62
for Thunderbird and from 0.69 to 0.68 for Log4j).

\begin{figure}[t]
  \centering
  \includegraphics[width=0.8\columnwidth]{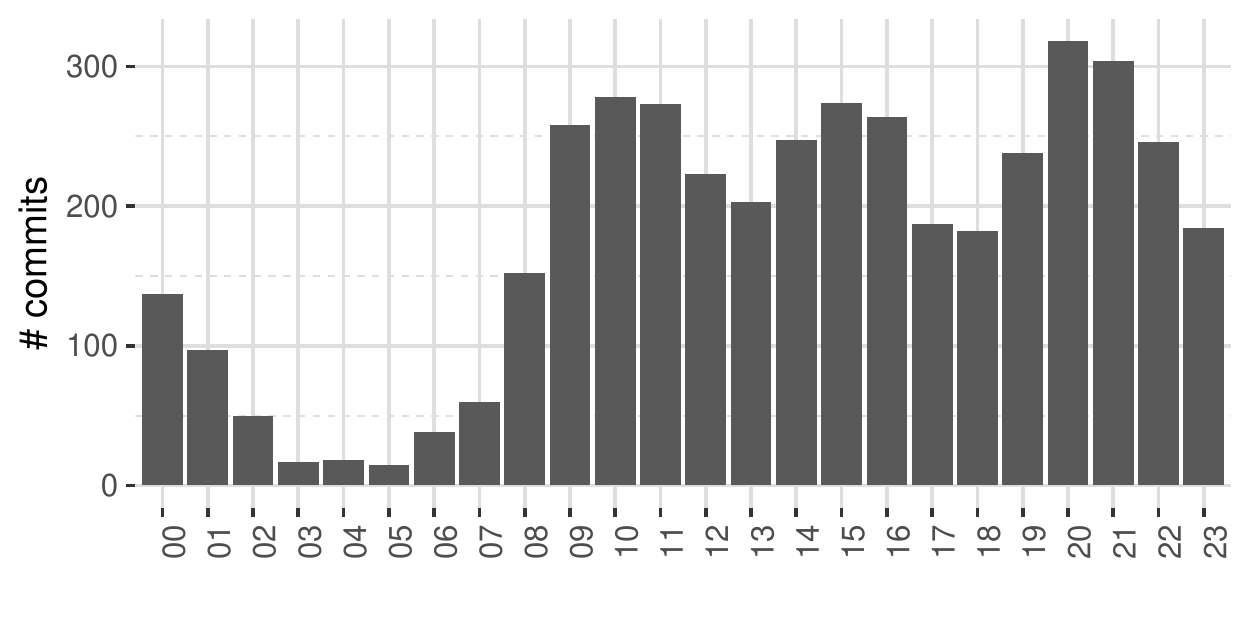}
  \caption{Distribution of the number of hourly weekday commits for Thunderbird.}
  \label{fig:thunderbird-hours}
\end{figure}

\textbf{Overall, we did not find any correlation between the amount
  of office hour work and shift of office hours period.} 
  Most of
the projects that experience large shifts have a rather high amount of activity outside the
identified office hour period. 
One possible explanation is that
commit activity is more spread throughout the day for these projects, 
making the identification of an 8-hour stint more sensitive to
small variations. For example, as seen in \fig{fig:thunderbird-hours},
Thunderbird has a high level of activity from 8 am to 1 am with
activity peaks around 10 am, 3 pm and 8 pm.

\textbf{Discussion:} In summary, most projects do not experience large changes over
  time in their work pattern. Important changes in the identified
  8-hour office hour period are often found in projects with commits
  being more spread throughout the day.


\section{Deeper analysis of the work patterns of Mozilla Firefox}\label{sec:analysis_ff}

\subsection*{RQ5. Are office hour commits different in terms of content?}

\begin{table}
\centering
\caption{Commit message differences (in \%) during vs. outside office hours.}
\label{tab:commit-diff}
\begin{tabular}{p{8em}|l|p{4em}|l}
Words, bi- or tri-grams   & Office hours & Outside office hours &
Difference \\
\hline
\hline
``backed out changeset''   & 3.65\%       & 2.63\%               & 138.7\%                \\
\hline
``back out''               & 0.57\%       & 1.24 \%                 & 46.0\%                 \\
\hline
``closed tree''            & 2.42\%       & 2.01\%               & 119.8\%                \\
\hline
add, adds, added         & 9.97\%       & 9.27\%               & 107.5\%                \\
\hline
remove, removes, removed & 7.43\%       & 8.16\%               & 91.1\%                 \\
\hline
fix, fixed, fixes        & 10.96\%      & 11.28\%              & 97.1\%
\end{tabular}
\end{table}

\begin{figure}[t]
  \centering
  \includegraphics[width=0.6\columnwidth]{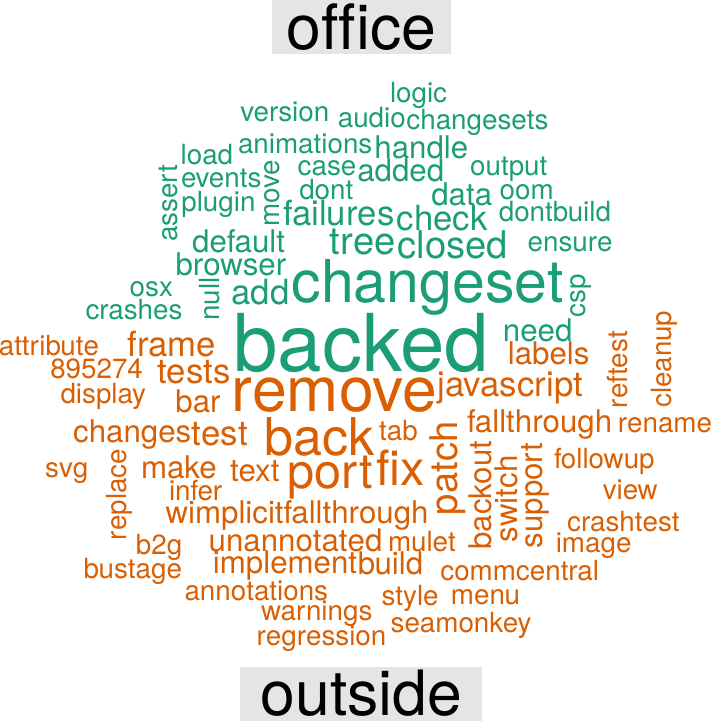}
  \caption{Comparison of the clouds of words used during and outside office hours.}
  \label{fig:comp_FF}
\end{figure}

\textbf{Motivation:} This question is similar to RQ3, which looked for differences between the
size of commits within and outside office hours, however here we investigate the content of the commits,
 by comparing word clouds. Figure~\ref{fig:comp_FF}
shows a comparison word cloud of commit messages for Mozilla Firefox, which has roughly 230,000 commits. As the visualization of a comparison
cloud tends to make the differences appear larger than they actually
are, \tab{tab:commit-diff} also shows percentage differences for the selected terms.

\textbf{Developers back out more code themselves outside office hours.} 
During office hours, there are more commit messages containing the ``Backed out
changeset'' statement, which is part of the official terminology of the
version control system used to indicate that a commit (change set) has been reverted (after being merged)
. Furthermore, ``Backed out changeset
<change set id>'' is often followed by the explanation such as ``because of a
possible Talos regression''\footnote{Talos is the performance testing framework
  used by Mozilla.}. For outside office hours there are more commit
messages with ``Back out'' (instead of ``Backed out changeset''), this informally seems to 
indicate that, after office hours, 
a developer is backing out changes herself rather than being backed out by some
authority.

A ``Closed tree''
message in a commit is used to indicate that a commit also closes a
version control tree. 
together with a ``Backed out changeset'' message. In fact, the ``closed
tree'' message is proportionally more common outside office hours when
we remove the commits where it appears together with the ``Backed out
changeset'' message (1.30\% vs 1.41\%). \tab{tab:commit-diff} also
shows that work involving adding something is more common during
office hours while removals and fixes are more common outside office
hours.


\textbf{Discussion:} We found small differences between office hour and outside-office hour commit messages. The most notable difference was that
  during office hours there are more formal reverts made to version
  control. Outside office hours, informal reverts made by the
  developer herself are more common.

\subsection*{RQ6. Can demographics explain office hour activity?}

\textbf{Motivation:} First, we hypothesize that more senior developers 
will likely work more during office hours, as they are more likely to be paid for their efforts. For
this purpose, we investigated whether there was a correlation between
the number of commits made by a developer and the amount of abnormal work. We only considered
individuals with 10 or more commits, which reduces the number of
analyzed individuals in Firefox significantly from 2,755 to 857, but still
retains 98\% of the commits.

Second, one explanation for work outside-office
hours in open source projects is the amount of paid and non-paid
contributors. If developers are hired by a company to work on a
project, they are more likely to work during regular office hours. On
the contrary, if they are not paid, they are more likely to work
during their free time. Hence, we ran a manual background check for the top 10\% of
developers to determine whether they are paid or non-paid, leaving us with 278 developers (out of 2,755), each with
147 or more commits. This top 10\% of developers has made 87\% out of all the
Firefox commits. Our background check considered information available online on
websites such as LinkedIn or the developers' personal websites (e.g.,
publicly available CVs). In addition to their job status, we also
gathered information about their location, experience in the software
industry and position at Mozilla.

\begin{figure}[t]
  \centering
  \includegraphics[width=0.8\columnwidth]{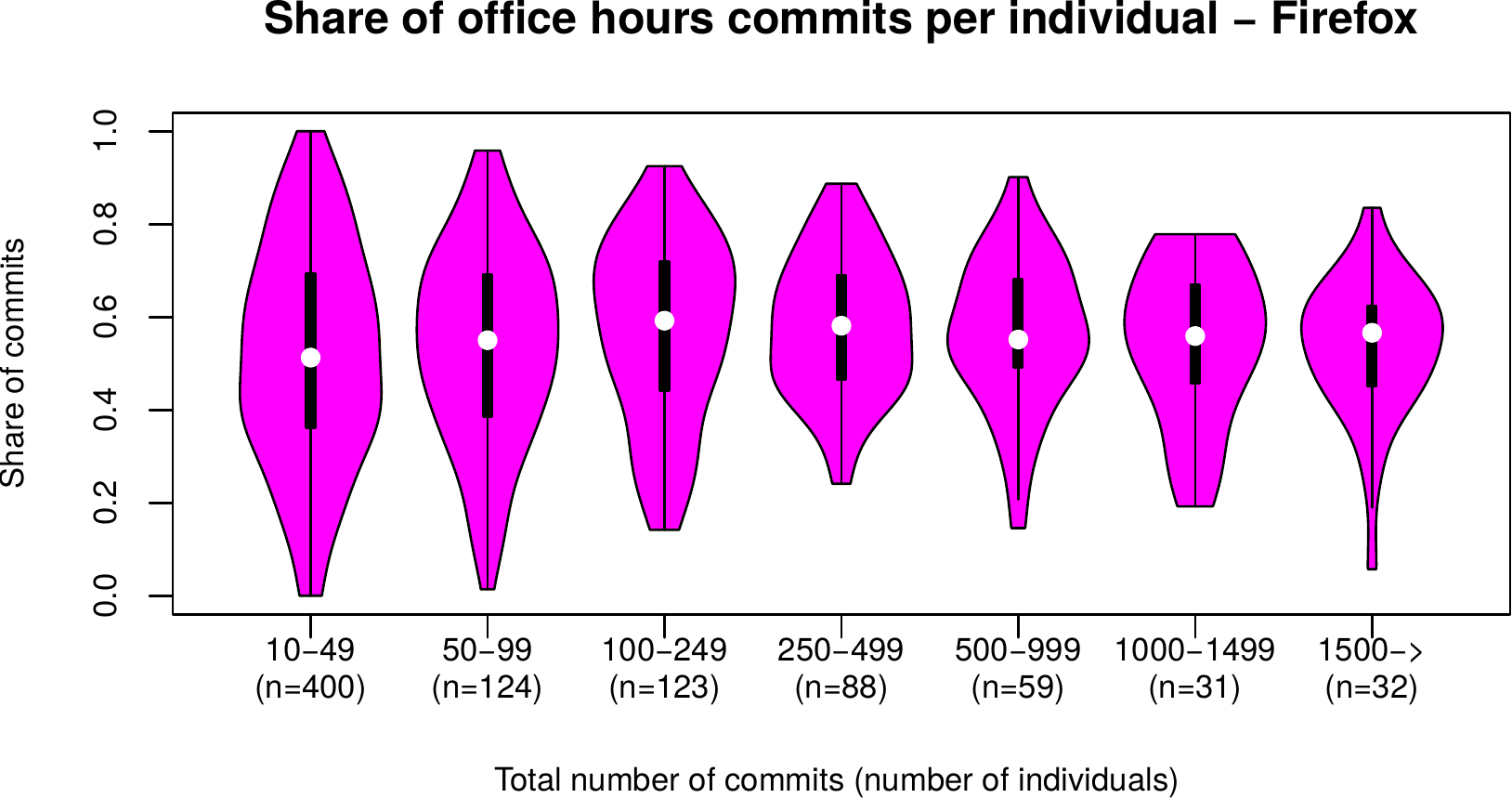}
  \caption{Share of office hour commits for individuals, for various numbers of total commits.}
  \label{fig:violin-mozilla}
\end{figure}

\textbf{We did not find a statistically significant
 correlation between developer seniority and amount of office hour work.} \fig{fig:violin-mozilla} divides the contributors into buckets \bram{what discretization approach was used? manual? equal distance? equal proportion?}
based on their number of commits. The violin plots show that the
amount of office hour work varies between groups. We can see an
increase in the amount of commits during office hours as the number of
commits increase in the first three groups. However, this increase
plateaus at the 4th cluster of developers (between 250-499
commits). Then, for the top contributor group with 1,500 or more
commits, the share of abnormal commits increases again. Overall, it
seems that the number of commits is not a very good predictor of office hours work.

\begin{figure}[t]
  \centering
  \includegraphics[width=0.8\columnwidth]{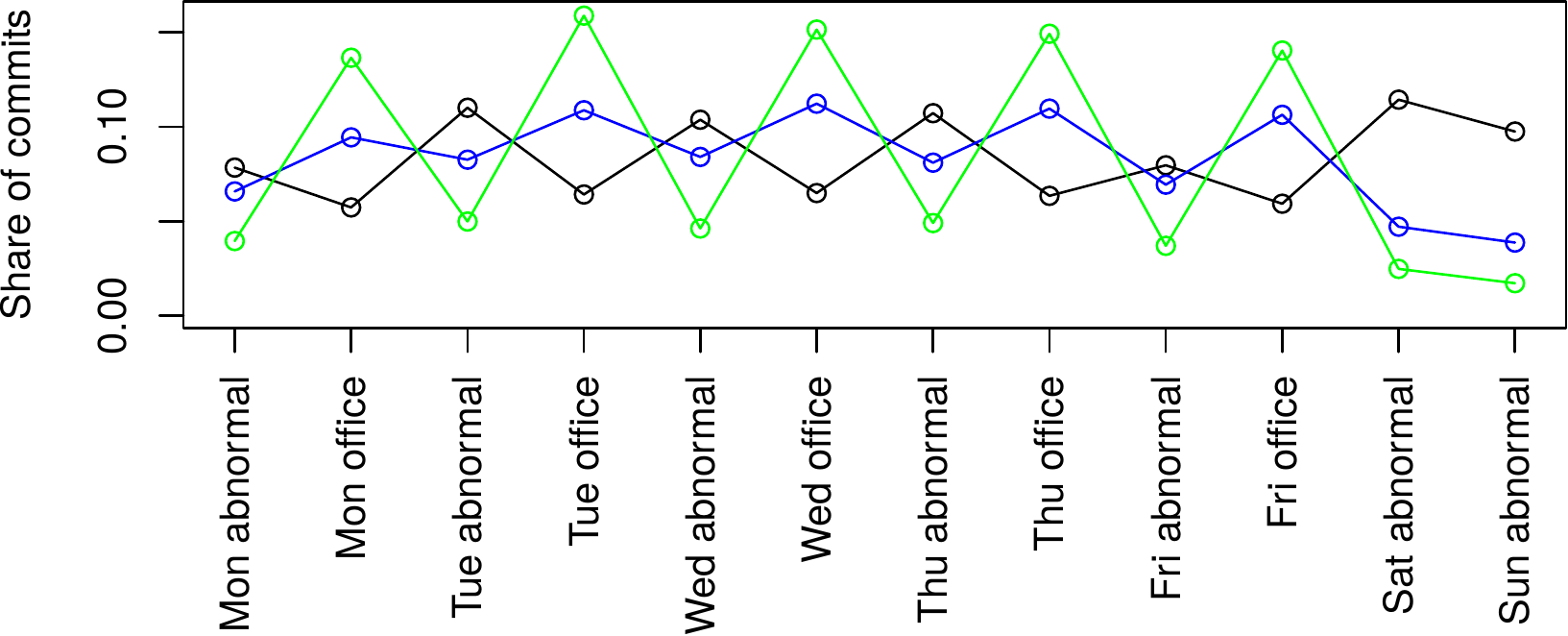}
  \caption{Clusters of work patterns for the top 10\% Firefox developers.}
  \label{fig:weekly-12period-mozilla}
\end{figure}

\textbf{Projects were clustered within 3 work pattern clusters, two of which mostly used office hours} 
\fig{fig:weekly-12period-mozilla} shows the clustering of three
different work patterns that were identified using k-means
clustering. Since we are interested in the differences between office
hour commits and outside-office hour commits within a single project,
we used our dynamic office hour setting data for Mozilla Firefox and
divided each week into 12 time units. From Monday through Friday, a
commit can happen either during office hours (normal) or
outside-office hours (abnormal), yielding 10 time units. For weekends,
we consider both Saturday and Sunday as abnormal, which gives two more
time units. Reducing the time units in a week (X-axis) from 168 hours
to 12 periods also makes the data less noisy and gives a visualization
that is easier to comprehend. Similar to
\fig{fig:All_weekly_hour_cluster}, \fig{fig:violin_commits}, and
\fig{fig:projects-2-extreme-clusters}, the Y-axis in
\fig{fig:weekly-12period-mozilla} represents the share of commits done
in a particular time
unit.

In \fig{fig:weekly-12period-mozilla}, we can see one group in cluster
(n=101) that mostly commits during office hours. This office hour
cluster has the highest share of commits (15.9\%) on Tuesday during
office hours and the lowest on Sunday (1.7\%). On the other hand, the
extreme (black) cluster (n=46) works more during abnormal times, and has
the highest share of commits on Saturday (11.4\%) and the lowest on
Monday during office hours (5.7\%). The third cluster in blue (n=131)
lies between these two extreme clusters and obtains the highest share
of commits on Wednesday during office hours (11.2\%) and the lowest on
Sunday (3.9\%). This group substantially works outside office hours
during the week, but less during weekends. For weekends, one needs to
consider that a single data point represents all the commits for the
day, while for a weekday the commits are divided into office hours and
abnormal commits.

For the three clusters in \fig{fig:weekly-12period-mozilla}, we ran
the dynamic office hour heuristic and found that the green and blue
clusters both have office hours from 10:00-18:00. During those hours
they perform 72 and 53\% of their weekly commits, respectively. The third cluster
 commits 34\% during their office hours, which are from 15:15 to
23:15, indicating a very irregular and abnormal working
pattern. Office hours for the entire Firefox project were 10:00-18:00,
and during that time the entire project completed 55\% of their
commits. Thus, the middle group represents the project average while
the other two groups show extreme variations.

\textbf{Outside office hour work is mostly performed by unpaid developers.} We compared the clusters identified in
\fig{fig:weekly-12period-mozilla} with our manually extracted
information about Mozilla paid developers. \tab{tab:cluster-hired}
shows the absolute number and relative percentage of paid developers in each
cluster (first two numbers in each cell). While 90\% of the developers in the cluster with
the largest amount of office hour commits (last column), and 88\% of the developers in the average cluster (second-to-last column)
were paid, only 50\% of the cluster with most outside office hours (second column) were paid by Mozilla. This corresponds to only 10.1\% of all paid developers. In other words, paid developers work less outside office hours than unpaid ones.
Yet most paid Firefox developers still work significantly more outside
office hours than developers from the local company

\begin{table}
  \centering
  \caption{Clustering of (un)paid top 10\% developers of Firefox. The first number and first percentage respectively indicate the number and percentage of developers in a given cluster that is (un)paid, while the second percentage indicates the percentage of (un)paid developers that are within a cluster.}
  \label{tab:cluster-hired}
  \begin{tabular}{l|p{7em}|p{7em}|p{7em}}
    Paid & Outside office hours cluster & Average cluster & Office hours cluster \\
    \hline
    No & 23 (50\% / 47.9\%) & 15 (11.6\% / 31.3\%)  & 10 (10\% / 20.8\%) \\
    Yes  & 23 (50\% / 10.1\%) & 114 (88.4\% / 50.2\%) & 90 (90\% / 39.7\%)
  \end{tabular}
\end{table}


\begin{table}
  \centering
  \caption{Median percentage of office hour commits depending on
    location, experience in the software industry and position at Mozilla.}
  \label{tab:abnormal-hired}
  \begin{tabular}{p{6em}|l|l|p{4em}|p{4em}}
    Characteristic & \# dev. & \# commits & office hours commits & beginning of office hours period \\
    \hline
    \textbf{Based in Europe}     & 65  & 41,962  & \textbf{54.4\%} & 9:45  \\
            Based in America     & 118 & 78,070  & 60.9\%          & 10:00 \\
    \hline
    \textbf{Senior position}     & 78  & 47,483  & \textbf{57.6\%} & 10:00 \\
            Non senior position  & 197 & 130,468 & 59.5\%          & 10:00 \\
    \hline
    \textbf{Manager position}    & 19  & 7,957   & \textbf{63.1\%} & 10:00 \\
            Non manager position & 256 & 169,994 & 58.6\%          & 9:30  \\
  \end{tabular}
\end{table}

\textbf{There are no major differences in office hour work between
  developer profiles.}  \tab{tab:abnormal-hired} shows that the 
developers based
in Europe work less during office hours. It seems that career moves on
the technical ladder, \ie having a title with the word ``senior'' or
``principle'', decreases the median office hour work by 1.9\%. However,
career moves on the managerial ladder, \ie having a title with the
word ``manager'', decrease the outside office hour commits by
4.5\%. We suspect that management positions require less technical
work and more communication and coordination, which might explain this
difference. In other words, for managers, using commits as working
hour indicators could be misleading. However, when running a
Mann–Whitney U test, only the difference between Europe-based and
America-based developers is statistically significant with a p-value
$< 0.01$.

\textbf{Discussion:} We did not find any differences between the
seniority of developers when measured in terms of commits. On the other hand,
\textbf{developers following office hours more closely are usually developers
paid by Mozilla}, even though most paid developers still work a lot
outside regular office hours. We did not observe any strong characteristics of paid
developers that could explain these differences. The only statistically
significant difference is observed between European and American
developers, with the former following office hours less than the latter. A
potential explanation is the difference in time zone and the fact that
Mozilla is based in North America. Finally, the office hour work
seems to be impacted by job title, but again the differences are not
statistically significant.



\section{Threats to Validity}\label{sec:threats}

Regarding construct validity, commit authoring times used in version control systems might not be
representative of the actual time period when the code change was
written: a timestamp cannot provide information about the time
actually spent working for that commit. However, as we investigate
weekly and daily rhythms, which are accumulated over months and years,
this should not significantly affect the results. For example, if a
developer works at a regular rhythm from nine to five, then, over time,
all commits would occur between these regular working hours, with
perhaps a small lag from the start of working until the first
commit. Furthermore, the fact that in most figures we observed dips in commit count
during typical lunch hours strengthens the idea that commit time
stamps are reliable when used for daily and weekly working hour
discovery.

However, for the local company, we investigated the
developers' chat log activity and commit activity. We found that not
only are these strongly correlated (r=0.74), but the ``start of the day
lag'' between the 8-hour office period based on the chat log
timestamps and the one based on the commit timestamps is only 15
minutes. This offers further support that commit time stamps can be
used to study what are the typical working hours in daily and weekly
level.

Another threat to construct valididty is that our analysis depends on data extracted from repositories available
from online sources. Although we found some missing time zone
information for Apache projects and filtered them out, other errors or
incompleteness in these data sources may impact the result of our
analysis. In particular, we relied on time zone
information available from Mozilla's Mercurial source code repository
and Apache source code repositories. The time zone from this
repository might not be accurate enough to pinpoint the developer's
actual position.

In order to identify developers hired by Mozilla, we manually looked
for information about them online. Although this allows us to identify a
large amount of the most active developers as hired by Mozilla, we
might have missed developers who do not share their CV online. We also merged developers' identities using a very basic identity merging
technique. We manually checked for false positives in order to avoid
merging the work pattern of two developers as a single one and thus
overestimating their amount of activity. However, there might be false
negatives remaining, which could be particularly problematic in the
case of developers using their work e-mail during office hours and
personal emails outside office hours.

Regarding threats to external validity, our study only includes open source projects from Mozilla and Apache,
and one project from a local company. The results obtained are
specific to those organizations' culture and to the habits of their developers. Although the results reveal important differences
between the projects, as illustrated in Figure~\ref{fig:projects-2-extreme-clusters}, we cannot guarantee that our
data set would be representative of the entire industry in particular,
as we would need more closed source projects.


\section{Conclusion and future work}\label{sec:conclusion}

In this paper, we have performed a first large-scale study of software
developers' work patterns by investigating over 700,000 version
control commits coming from 87 software projects, each having a minimum of
2,000 commits. We were
motivated to investigate work patterns as according to medical and
occupation well-being literature they might be possible indicators of
overload, time pressure, and unhealthy working patterns.

We found that developers follow a typical circadian rhythm where most
of the work is performed during day time and less work is performed
during evenings and at night. Dips in activity during typical lunch
hours can also be seen. The weekly rhythm shows that during weekends
there is radically less activity. We then established a search method
that finds, for each project or individual, the period, of a given
length, with the highest number of commits. In this paper, we fixed
the length of time series to eight hours to make comparisons between
projects. Our method can also be used in reverse to find a continuous
time series that results in a certain percentage of commits.

We used our search method to find that the median 8-hour working day
for software engineering projects in our data runs from 10:00 to 18:00
instead of the typical 09:00-17:00 (nine-to-five). Thus, the often
quoted idea of programmers as night owls that has even made it to a
book titled ``Why programmers work at night''~\cite{Teller2013} holds
only partially. Perhaps the plus one hour shift compared to regular
nine-to-five professions gives the impression of a highly elevated
night activity. Another possibility for the birth of that myth could
be in extreme projects as the ones shown in
Figure~\ref{fig:projects-2-extreme-clusters}, with a blue line showing
that a peak hour of commits occurs daily very close to midnight.

In fact, a previous comparison of the work rhythms of software
engineers and scientists~\cite{DBLP:journals/corr/abs-1208-2686} shows
that scientists seem to work more during the night than software
engineers. The activity of US and German scientists at night (measured
by paper downloads) is roughly 42\% and 25\% of their day time peak
activity. On the other hand, for software developers in the Mozilla
and Apache projects, the corresponding percentages (based on commit count)
are roughly 29\% and 25\% (see \fig{fig:apache-hours} and
\fig{fig:mozilla-hours}).

Of course, individuals have different working patterns and, indeed,
clustering of developers revealed that one third of software
developers do not seem to follow a typical office hour rhythm at all
as they perform a lot of activities during evenings and weekends (see
Figure~\ref{fig:All_weekly_hour_cluster}). On the other hand, two
thirds of individuals mostly follow office hours. This information is
important if one wishes to build a stress detector that would use
among other things commit hours, since such a detector would only work
for the two thirds of software engineers who have regular working
rhythms.

A manual background check of the top 10\% (over 250 individuals) of
Mozilla Firefox developers revealed that only half of the developers
that did not follow office hours were paid by Mozilla (black line in
Figure~\ref{fig:weekly-12period-mozilla}) as opposed to roughly 90\%
for the two groups that did follow office hours. This suggests that
existence of hobby developers can partially explain the outside office
hour activity. Still, for the middle group (blue curve in
Figure~\ref{fig:weekly-12period-mozilla}), only 55\% of commits are
performed during office hours, although 90\% of the individuals are
paid employees, which could be a marker of overload for that group.


With respect to seniority we found no differences in working hours by using two different measures the number of commits and job title. 
This is rather surprising but perhaps developers that make very few commits are
hobbyist and commit outside office hours while for the most senior developers the outside office hour commits would be explained by overwork.

A comparison of commit messages between office and outside office
hours revealed that in our largest project, Firefox, formal code
review reverts are made more often during office hours. Moreover,
informal developer-initiated reverts are more often made outside
office hours. Furthermore, office hour time seems to focus especially
on adding new code while clean-ups are more frequent outside office
hours. We found no difference in commit size, and we did not find
support to the idea that as projects mature they move away from night
and weekend work to more typical office hour rhythm.

In future work, we will perform a detailed qualitative analysis of
both individual histories and individual developer histories to better
explain the differences in working patterns, involving interviews and
daily surveys to monitor project members. We will use additional data
sources, such as chat logs, in order to make our set of timestamp
activity more complete. Measuring the level of detachment from work,
which is critical to recovery, should be studied and, similarly,
normative recommendation from empirical data should be drawn. We also
intend to study the impact of policies and guidelines put in place by
project managers, such as a fast release cycle, on developers'
activity and health. Finally, we also want to investigate further the
causes of outside office hour work by focusing on periods with
unusually high activity during the night or weekends.



\section*{Acknowledgments}

The first three authors have been supported by Academy of
Finland grant 298020. The third author has been supported by Kaute foundation. 





\bibliographystyle{IEEEtran}
\bibliography{biblio}

\end{document}